\RequirePackage{ifpdf}

\RequirePackage{fixltx2e}
\RequirePackage{fix-cm}

\documentclass[%
floatfix,
showkeys,
nofootinbib, %
superscriptaddress, %
]{revtex4-2}

\usepackage{iftex}
\ifpdftex
\usepackage{mmap}		
\fi

\usepackage{textcomp}

\usepackage[T1,T2A]{fontenc}

\usepackage[utf8]{inputenx}
\input{ix-utf8enc.dfu}

\usepackage[german,russian,english]{babel}

\usepackage{amsmath}
\usepackage{amssymb}

\usepackage{mathtools}
\mathtoolsset{
showonlyrefs,
mathic = true
}

\allowdisplaybreaks

\usepackage{hyperref}
\hypersetup{backref,
 colorlinks=false}
\hypersetup{pdfborder=0 0 0}

\usepackage{microtype}
\UseMicrotypeSet[protrusion]{alltext}

\usepackage{graphicx}

\usepackage[scanall]{psfrag}

\usepackage{listings}
\usepackage{listingsutf8}
\usepackage[frozencache]{minted}

\setminted{breaklines=true}

\usepackage{fixltx2e}

\usepackage{nicefrac}

\makeatletter
\def\ps@pprintTitle{%
     \let\@oddhead\@empty
     \let\@evenhead\@empty
     \let\@oddfoot\@empty
     \let\@evenfoot\@oddfoot}
\makeatother

\begin{document}

\graphicspath{{image/multiple-dispatch/en/}{image/multiple-dispatch/}{image/}}

\title{Implementation of hyperbolic complex numbers in Julia language}

\author{Anna V. Korolkova}
\email{korolkova-av@rudn.ru}
\affiliation{
  Peoples' Friendship University of Russia (RUDN University),\\
  6 Miklukho-Maklaya St, Moscow, 117198, Russian Federation}

\author{Migran N. Gevorkyan}
\email{gevorkyan-mn@rudn.ru}
\affiliation{
  Peoples' Friendship University of Russia (RUDN University),\\
  6 Miklukho-Maklaya St, Moscow, 117198, Russian Federation}

\author{Dmitry S. Kulyabov}
\email{kulyabov-ds@rudn.ru}
\affiliation{
  Peoples' Friendship University of Russia (RUDN University),\\
  6 Miklukho-Maklaya St, Moscow, 117198, Russian Federation}
\affiliation{
  Joint Institute for Nuclear Research\\
  6 Joliot-Curie, Dubna, Moscow region, 141980, Russian Federation}

\begin{abstract}
\textbf{Background:}
Hyperbolic complex numbers are used in the description of hyperbolic
spaces. One of the well-known examples of such spaces is the Minkowski
space, which plays a leading role in the problems of the special
theory of relativity and electrodynamics. However, such numbers are
not very common in different programming languages.
\textbf{Purpose:}
Of interest is the implementation of hyperbolic complex in scientific
programming languages, in particular, in the Julia language.
\textbf{Methods:}
The Julia language is based on the concept of multiple dispatch. This
concept is an extension of the concept of polymorphism for
object-oriented programming languages. To implement hyperbolic complex
numbers, the multiple dispatching approach of the Julia language was
used.
\textbf{Results:}
The result is a library that implements hyperbolic numbers.
\textbf{Conclusions:}
Based on the results of the study, we can conclude that the concept of
multiple dispatching in scientific programming languages is convenient
and natural.
\end{abstract}


\maketitle

  \section{Introduction}

The Julia programming language~\cite{bezanson_2017_julia-numeric-computing,
  kulyabov_2014_conf_grid_julia} is a promising language for scientific computing. At the moment, the Julia language has reached a stable state. By design, Julia solves the \emph{problem of two languages}. This problem lies in the fact that for rapid prototyping, data processing and visualization, an interpreted dynamic language or a mathematical package (Python, Matlab, etc.) is used, and for intensive numerical calculations, the program has to be rewritten in a compiled language with static typing (C/ C++, Fortran).

An illustration of this problem can be seen in
Python, which has gained wide
popularity as an interface \emph{language-glue}. Numerous wrapper
libraries were written on it, which used Python code to call C/C++ and
Fortran functions from precompiled libraries. For example,
the well-known library NumPy~\cite{oliphant_book_guide-numpy_en} consists of 51\% C code
and only 47\% Python code (the remaining percentages are divided
between C++, Fortran, JavaScript and Unix
shell).

The Julia language combines the flexibility of dynamically typed interpreted languages with the performance of statically typed compiled languages.

The basic part of the Julia language is very similar to other
scientific programming languages, so it does not cause difficulties in
mastering. However, Julia's core is built around the concept of \emph{multiple
  dispatch}~\cite{zappa-nardelli_2018_julia-subtyping_en}, which is
rare in other languages. It is in this mechanism that the essential
difference of Julia from other languages lies, and its understanding
is essential for the full use of all the advantages of Julia.

\subsection{Paper structure}
\label{sec:structure}

In the article, the authors paid great attention to illustrating the
mechanism of multiple dispatch and other mechanisms that are closely
related to it.

In the first part of the article, we give the necessary definitions
and illustrate the concept of multiple dispatch with simple examples
that allow you to understand the syntax associated with this part of
the language and capture the essence of this approach. In the second
part, we give an example of the implementation of hyperbolic complex
numbers in the Julia language. This example allows you to touch not
only multiple dispatch, but also the type casting mechanism, the
abstract type hierarchy, overloading arithmetic operators, and
specifying user-defined data types.

  \section{Multiple dispatch}

  \subsection{Common definitions}

\emph{Dynamic dispatch} is a mechanism that allows you to choose which
of the many implementations of a polymorphic function (or operator)
should be called in a given case~\cite{driesen_1995_message-dispatch_en}. In this case, the choice of one or
another implementation is carried out at the stage of program
execution. \emph{Multiple dispatch} is based on dynamic dispatch. In
this case, the choice of implementation of a polymorphic function is
made based on the type, number, and order of the function's
arguments. This is how runtime polymorphic dispatch is
implemented~\cite{muschevici_2008_multiple-dispatch-practice_en, gowda_2022_symbolic-numerics-multiple-dispatch_en}. Note also that in addition to the term multiple dispatch,
the term \emph{multimethod} is also used.

The mechanism of multiple dispatch is similar to the mechanism of
overloading functions and operators, implemented, for example, in the
C++ language. Function overloading, however, is done exclusively at
compile time, while multiple dispatch should work at runtime as well
(runtime polymorphism).

  \subsection{Multiple dispatch in Julia}

To illustrate the mechanism of multiple dispatch, we will give the
following code example in the Julia language.

\begin{minted}{julia}
function f(x, y)
  println("Generic implementation")
  return x + y
end

function f(x)
  println("For single argument")
  return x
end

function f(x::Integer, y::Integer)
  println("Implementation for integers")
  return x + y
end

function f(x::String, y::String)
  println("Implementation for strings")
  return x * " " * y
end

function f(x::Tuple{Int, Int}, y::Tuple{Int, Int})
  println("Implementation for tuples of two integer elements")
  return (x[1], x[2], y[1], y[2])
end
\end{minted}

In this example, we have created five implementations of the $f$
function, which differ from each other in different signatures. In
terms of the Julia language, this means that one function $f$ now has
four different \emph{methods}. In the first two methods, we did not
use type annotations, so the type of the arguments will be determined
either at compile time or at run time (as in interpreted
languages). It is also worth noting that Julia uses dynamic JIT
compilation (just-in-time), so the compilation stage is not explicitly
separated from the execution stage for the user.

The arguments of the following three methods are annotated with types,
so they will only be called if the types match the annotations. In the
\verb|f| for strings, the \texttt{*} concatenation operator is
used. The choice of the multiplication sign \texttt{*} instead of the
more traditional addition sign \texttt{+} is justified by the creators
of the language by the fact that string concatenation is not a
commuting operation, so it is more logical to use the multiplication
sign for it, rather than the addition sign, which is often used to
denote commuting operations.

The following code snippet illustrates how multiple dispatch works at
compile time. The \mintinline{julia}{@show} macro is used to print out
the name of the function and the arguments passed to it.
\begin{minted}{julia}
  @show f(2.0, 1)
  @show f(2, 2)
  @show f(0x2, 0x1) # numbers in hexadecimal system
  @show f("Text", "line")
  @show f(3)
  @show f([1, 2], [3, 4])
  @show f((1, 2), (3, 4))
\end{minted}

  \begin{itemize}
  \item In the first line, we passed real (floating-point) type
    arguments to the function, so a generic implementation call was
    made. Since the operator \verb|+| is defined for floating point
    numbers, the function succeeded and gave the correct result.
  \item Methods for integers were called in the second and third
    lines. Note that the \mintinline{julia}{Integer} type is an
    \emph{abstract} type and includes signed and unsigned integers
    from 1 to 16 bytes in size, defined in the language core. Numbers
    written in hexadecimal are interpreted by default as unsigned
    integers.
  \item The method for strings was called on the fourth line. In the
    fifth line, the method for one argument.
  \item The sixth line passed two arrays as arguments. The \texttt{+}
    operation is defined for arrays, so the function ran without error
    and returned an element-wise sum.
  \item In the seventh line, the function arguments are tuples
    consisting of two integers. Since we defined a method for such a
    combination of arguments, the function worked correctly.
  \end{itemize}

\begin{minted}{julia}
  Generic implementation
  f(2.0, 1) = 3.0
  Implementation for integers
  f(2, 2) = 4
  Implementation for integers
  f(0x02, 0x01) = 0x03
  Implementation for strings
  f("Text", "line") = "Text line"
  For single argument
  f(3) = 3
  Generic implementation
  f([1, 2], [3, 4]) = [4, 6]
  Implementation for tuples of two integer elements
  f((1, 2), (3, 4)) = (1, 2, 3, 4)
\end{minted}

The above examples will work correctly in languages that support
function overloading and do not demonstrate the specifics of dynamic
dispatching, since the types of arguments are known at the compilation
stage and are available to the translator.

To test the work of dynamic method calls, consider the following code:
\begin{minted}{julia}
  print("Enter an integer:")
  # Read a string and convert to an integer type
  @show n = parse(Int32, readline())
  if n > 0
    x = 1.2; y = 0.1
  else
    x = 1; y = 2
  end
f(x, y)
\end{minted}
Here, the types of variable values \verb|x| and \verb|y| are not known
at compile time, as they depend on what number the user enters during
program execution. However, for the case of integer \verb|x| and
\verb|y| the corresponding method is called.

  \section {Hyperbolic numbers}

We will use hyperbolic numbers to illustrate the multiple dispatch
capabilities of the Julia language, so we will limit ourselves to the
definition and basic arithmetic operations.

\emph{Hyperbolic numbers}~\cite{yaglom_complex-numbers-geometry_en, yaglom_1964_projective-metrics_en, kulyabov_2021_preprint_complex-numbers-relativistic-operations_en, kulyabov_2020_jopcs_hyperbolic-numbers_en},
along with elliptic and parabolic numbers, are a generalization of
complex numbers. Hyperbolic numbers can be defined as follows:
\begin{equation}
  z = x + \mathrm{j} y,\; \mathrm{j}^2 = 1,\; \mathrm{j}\neq \pm 1.
\end{equation}
The quantity $j$ will be called the \emph{hyperbolic imaginary unit},
and the quantities $x$ and $y$ will be called the real and imaginary
parts, respectively.

For two hyperbolic numbers $z_1 = x_1 + \mathrm{j}y_1$ and $z_2 = x_2
+ \mathrm{j}y_2$ the following arithmetic operations are performed.
\begin{description}
  \item [Addition] $z_1 + z_2 = (x_1 + x_2) + \mathrm{j}(y_1 + y_2)$.
  \item [Multiplication] $z_1 z_2 = (x_1 x_2 + y_1 y_2) + \mathrm{j}(x_1 y_2 + x_2 y_1)$.
  \item [Conjugation] $z^{\ast} = x - \mathrm{j} y$.
  \item [Inverse number] $z^{-1} = \dfrac{x}{x^2 + y^2} - \mathrm{j}\dfrac{y}{x^2 - y^2}$.
  \item [Division] $\dfrac{z_1}{z_2} = \dfrac{x_1 x_2 - y_1 y_2}{x^2_2 - y^2_2} + \mathrm{j}\dfrac{x_1 y_1 - x_1 y_2}{x^2_2 - y^2_2}$.
\end{description}

The implementation of hyperbolic numbers is in many respects similar
to the implementation of complex ones. Operators \verb|+|, \verb|-|,
\verb|*| must be overloaded, and \verb|/|, root extraction,
exponentiation, elementary math functions, etc. At the same time, for
the purposes of illustrating the mechanism of operation of multiple
dispatching, it is arithmetic operations that are of primary
interest. This is due to the fact that elementary functions take only
one argument, and it is enough to define only one method for them. In
the case of arithmetic operators, it is necessary to provide
combinations of arguments of different numeric types. So, for example,
it should be possible to add a hyperbolic number to an integer,
rational, irrational number, which automatically affects not only
multiple dispatch, but also type casting mechanisms, an abstract type
hierarchy, and default constructor overloading.

Therefore, we will confine ourselves to examples of the implementation
of precisely arithmetic operations and that’s all, without touching on
the more mathematically complex calculations of various elementary
functions of a hyperbolic number.

Note that in addition to the term hyperbolic numbers, there are also
terms in the literature: double numbers, split complex numbers,
perplex numbers, hyperbolic
numbers~\cite{fjelstad_1986_extending-sr-perplex-numbers_en,
  band_1988_sr-perplex-numbers-comments_en,
  yaglom_complex-numbers-geometry_en, rooney_1978_three-types-complex-number_en, rooney_2014_generalised-complex-numbers_en}.

  \section{Implementation of hyperbolic numbers in Julia}

  \subsection{Declaring a Data Structure}

\begin{figure}
  \centering
  \includegraphics[width=0.8\linewidth]{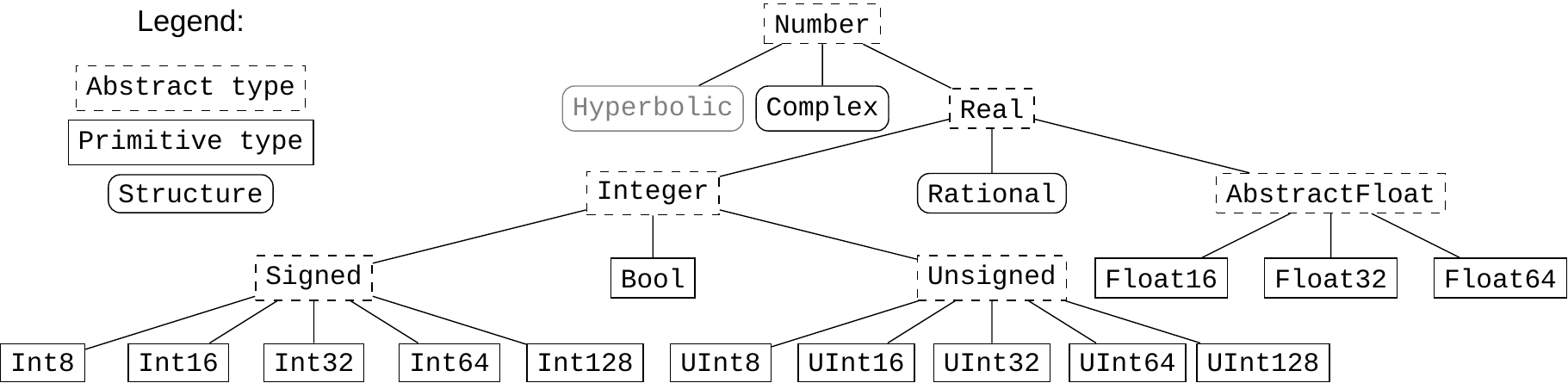}
  \caption{Location of Hyperbolic Numbers in Julia's Type Hierarchy}
  \label{fig:01}
\end{figure}

The implementation of hyperbolic numbers in Julia was based on the
code for complex numbers available in the official Julia
repository.
We also used the developments obtained in the implementation of parabolic complex numbers~\cite{kulyabov_2020_ceur-ws_2639_dual-complex_en}.
New type \verb|Hyperbolic|
defined with an immutable structure:
\begin{minted}{julia}
  struct Hyperbolic{T<:Real} <: Number
    "Real part"
    re::T
    "Imaginary part"
    jm::T
  end
\end{minted}

The structure is simple and contains only two fields of parametric
type \verb|T|. This requires that the type \verb|T| was a subtype of
the abstract type \mintinline{julia}|Real| (syntax
\mintinline{julia}|T<:Real|). The type \verb|Hyperbolic| is a subtype
of the abstract type \mintinline{julia}|Number| (see Fig.~\ref{fig:01}). Thus, hyperbolic
numbers are built into an already existing hierarchy of numeric types.

After the structure is defined, a new object of type \verb|Hyperbolic|
can be created by calling the default constructor. So, for example,
the number $h = 1 + \mathrm{j}3$ is given as follows:
\begin{minted}{julia}
  h = Hyperbolic{Float64}(1, 3)
\end{minted}

After creation, you can access the fields of the structure as
\mintinline{julia}{h.re} and \mintinline{julia}{h.jm}, but an attempt
changing the value of a field of an already existing object will
result in an error, since structs are immutable entities.

\begin{minted}{julia}
h = Hyperbolic(1, 3)
\end{minted}

However, if the argument types are different, then the default
constructor will not be able to implicitly cast and create a new
object. In this case, you must explicitly specify the parametric type
\begin{minted}{julia}
  # Float64 и Int64
  h = Hyperbolic(1.0, 3) # Error
  h = Hyperbolic{Float64}(1.0, 3) # Correct
\end{minted}

  \subsection{Additional constructors}

The default constructor is a normal function whose name is the same as
the type name. By creating additional methods for this function, you
can create additional constructors to handle various special cases.

So, for example, in order not to specify a parametric type every time,
you should add a new constructor of the following form:
\begin{minted}{julia}
  """Constructor №2"""
  function Hyperbolic(x::Real, y::Real)
    return Hyperbolic(promote(x, y)...)
  end
\end{minted}

The \mintinline{julia}{promote} function casts the arguments passed to
it to a common type and returns the result as a tuple. Postfix
operator \verb|...| unpacks the tuple and passes its elements as
arguments to the constructor function. The language core defines
casting rules for all subtypes of the \mintinline{julia}|Real|
abstract type, so now the constructor will work correctly for any
combination of arguments, as long as the \mintinline{julia}|T<:Real|
rule is fulfilled. For example, the following code will work
correctly:
\begin{minted}{julia}
  # Rational и Float64
  h = Hyperbolic(1//3, pi)
  >> Hyperbolic{Float64}(0.5, 3.141592653589793)
\end{minted}

We passed a rational number (type \mintinline{julia}|Rational|) and a
built-in global constant (number $\pi$) of type
\mintinline{julia}|Float64| to the constructor. After that, the type
casting rule worked and both arguments were cast to the type
\mintinline{julia}|Float64| as more general.

Declaring two more additional constructors will allow you to specify
hyperbolic numbers with zero imaginary part:
\begin{minted}{julia}
  """Constructor №3"""
  function Hyperbolic{T}(x::Real) where {T<:Real}
    return Hyperbolic{T}(x, 0)
  end
  """Constructor №4"""
  function Hyperbolic(x::Real)
    return Hyperbolic(promote(x, 0)...)
  end
\end{minted}

Constructor number 3 is a parametric function that is declared using
the \mintinline{julia}|where| construct. The \verb|T| is a subtype of
the abstract type \mintinline{julia}|Real|. Constructor number 4 works
similarly to constructor number 2.

Two more constructors will allow you to pass other hyperbolic numbers
as an argument to the constructor.
\begin{minted}{julia}
  """Constructor №5"""
  function Hyperbolic{T}(h::Hyperbolic) where {T<:Real}
    Hyperbolic{T}(h.re, h.jm)
  end
  """Constructor №6"""
  function Hyperbolic(h::Hyperbolic)
    return Hyperbolic(promote(h.re, h.jm)...)
  end
\end{minted}

For more convenience, you can also create a separate constant for
the imaginary cost $\mathrm{j}$:
\begin{minted}{julia}
  const jm = Hyperbolic(0, 1)
\end{minted}

  \subsection{Data printing}

To be able to print hyperbolic type values in a compact and readable
form, you should add the appropriate methods to the
\mintinline{julia}|show| function from the \mintinline{julia}|Base|
module.
\begin{minted}{julia}
  function Base.show(io::IO, h::Hyperbolic)
    print(io, h.re, "+", h.jm, "j")
  end
\end{minted}

Function \mintinline{julia}|show| is used when printing data to the
console, in particular, it is called by the
\mintinline{julia}|println| and macro \mintinline{julia}|@show|. The
code and output listings below will assume that the
\mintinline{julia}|show| method has been added  for hyperbolic
numbers.

  \subsection{Type casting}

Before proceeding to the implementation of methods for arithmetic
operations with hyperbolic numbers, it is necessary to define the
rules for type casting. To do this, create a new method for the
function \mintinline{julia}|promote_rule| from the
\mintinline{julia}|Base| module.
\begin{minted}{julia}
  function Base.promote_rule(::Type{Hyperbolic{T}}, ::Type{S}) where {T<:Real, S<:Real}
    return Hyperbolic{promote_type(T, S)}
  end
  function Base.promote_rule(::Type{Hyperbolic{T}}, ::Type{Hyperbolic{S}}) where {T<:Real, S<:Real}
    return Hyperbolic{promote_type(T, S)}
  end
\end{minted}

As arguments in \mintinline{julia}|promote_rule| parametric types are
specified, which should be cast to one enclosing type. In our case,
this is possible if one of the types is a subtype of
\mintinline{julia}|Real|, then the enclosing type is
\mintinline{julia}|Hyperbolic|.

After adding methods for \mintinline{julia}|promote_rule|, it becomes
possible to use functions \mintinline{julia}|promote|,
\mintinline{julia}|promote_type| and \mintinline{julia}|convert|.
\begin{minted}{julia}
  >>h = Hyperbolic(1 // 2)
  >>promote(h, 1)
  (1//2+0//1j, 1//1+0//1j)
  >>promote_type(Hyperbolic{Int64}, Float32)
  Hyperbolic{Float32}
\end{minted}

The first function is already familiar to us. The second allows you to
infer the enclosing type not of specific variable values, but of the
types themselves. A type in Julia is an object of the first kind (type
\mintinline{julia}|DataType|) and can be assigned to other variables,
passed as function arguments, and so on.

Function \mintinline{julia}|convert| allows you to convert the type
specific value, for example:
\begin{minted}{julia}
  >>convert(Hyperbolic, 1)
  1+0j
\end{minted}

After adding methods for type casting, you can start adding methods
for arithmetic operations. A feature of Julia is the implementation of
arithmetic operations not in the form of operators, but in the form of
functions. For example, the following calls are correct:
\begin{minted}{julia}
>>+(1,2)
3
>>+(1,2,3,4)
10
>>+((i for i in 1:10)...) # числа от 1 до 10
55
\end{minted}

In this regard, adding methods for arithmetic operations is no
different from the corresponding process for other functions.

Adding methods for unary operations \verb|+| and \verb|-| is carried out as follows:
 \begin{minted}{julia}
  Base.:+(h::Hyperbolic) = Hyperbolic(+h.re, +h.jm)
  Base.:-(h::Hyperbolic) = Hyperbolic(-h.re, -h.jm)
\end{minted}
This is an abbreviated function declaration.  

Similarly, methods are added for binary addition, subtraction,
multiplication, and division. Here is the code for addition and
multiplication.
\begin{minted}{julia}
  # Binary + and *
  function Base.:+(x::Hyperbolic, y::Hyperbolic)
    xx = x.re + y.re
    yy = x.jm + y.jm
    Hyperbolic(xx, yy)
  end
  function Base.:*(x::Hyperbolic, y::Hyperbolic)
    xx = x.re * y.re + x.jm * y.jm
    yy = x.re * y.jm + x.je * y.re
    return Hyperbolic(xx, yy)
  end
\end{minted}

    \section{Conclusion}

We examined the mechanism of multiple dispatch underlying the Julia
language, using the example of the implementation of hyperbolic
numbers. This example allowed us to touch upon such concepts of the
language as the hierarchy of data types, composite data types, type
casting mechanisms, function overloading (creating new methods for
functions in terms of the Julia language), etc.

\begin{acknowledgments}

This paper has been supported by the RUDN University Strategic Academic Leadership Program.

\end{acknowledgments}

 \bibliographystyle{elsarticle-num}

\bibliography{bib/multiple-dispatch/cite}

\begin{thebibliography}{10}
\expandafter\ifx\csname url\endcsname\relax
  \def\url#1{\texttt{#1}}\fi
\expandafter\ifx\csname urlprefix\endcsname\relax\def\urlprefix{URL }\fi
\expandafter\ifx\csname href\endcsname\relax
  \def\href#1#2{#2} \def\path#1{#1}\fi

\bibitem{bezanson_2017_julia-numeric-computing}
J.~Bezanson, A.~Edelman, S.~Karpinski, V.~B. Shah, Julia: A fresh approach to
  numerical computing, SIAM Review 59~(1) (2017) 65--98.
\newblock \href {https://doi.org/10.1137/141000671}
  {\path{doi:10.1137/141000671}}.

\bibitem{kulyabov_2014_conf_grid_julia}
M.~N. Gevorkyan, D.~S. Kulyabov, L.~A. Sevastyanov, Review of julia programming
  language for scientific computing, in: The 6th International Conference
  "Distributed Computing and Grid-technologies in Science and Education", 2014,
  p.~27.

\bibitem{oliphant_book_guide-numpy_en}
T.~E. Oliphant, Guide to NumPy, 2nd Edition, CreateSpace Independent Publishing
  Platform, 2015.

\bibitem{zappa-nardelli_2018_julia-subtyping_en}
F.~{Zappa Nardelli}, J.~Belyakova, A.~Pelenitsyn, B.~Chung, J.~Bezanson,
  J.~Vitek, Julia subtyping: a rational reconstruction, Proceedings of the ACM
  on Programming Languages 2~(OOPSLA) (2018) 1--27.
\newblock \href {https://doi.org/10.1145/3276483} {\path{doi:10.1145/3276483}}.

\bibitem{driesen_1995_message-dispatch_en}
K.~Driesen, U.~Hölzle, J.~Vitek, Message Dispatch on Pipelined Processors,
  Lecture Notes in Computer Science, Springer Berlin Heidelberg, 1995.
\newblock \href {https://doi.org/10.1007/3-540-49538-x_13}
  {\path{doi:10.1007/3-540-49538-x_13}}.

\bibitem{muschevici_2008_multiple-dispatch-practice_en}
R.~Muschevici, A.~Potanin, E.~Tempero, J.~Noble, Multiple dispatch in practice,
  in: OOPSLA'08: Proceedings of the 23rd ACM SIGPLAN conference on
  Object-oriented programming systems languages and applications, ACM Press,
  2008, p. 563–582.
\newblock \href {https://doi.org/10.1145/1449764.1449808}
  {\path{doi:10.1145/1449764.1449808}}.

\bibitem{gowda_2022_symbolic-numerics-multiple-dispatch_en}
S.~Gowda, Y.~Ma, A.~Cheli, M.~Gwóźzdź, V.~B. Shah, A.~Edelman,
  C.~Rackauckas, High-performance symbolic-numerics via multiple dispatch, ACM
  Communications in Computer Algebra 55~(3) (2022) 92–96.
\newblock \href {https://doi.org/10.1145/3511528.3511535}
  {\path{doi:10.1145/3511528.3511535}}.

\bibitem{yaglom_complex-numbers-geometry_en}
I.~M. Yaglom, Complex Numbers in Geometry, Academic Press, 1968.

\bibitem{yaglom_1964_projective-metrics_en}
I.~M. Yaglom, B.~A. Rozenfel'd, E.~U. Yasinskaya, Projective metrics, Russian
  Mathematical Surveys 19~(5) (1964) 49--107.
\newblock \href {https://doi.org/10.1070/RM1964v019n05ABEH001159}
  {\path{doi:10.1070/RM1964v019n05ABEH001159}}.

\bibitem{kulyabov_2021_preprint_complex-numbers-relativistic-operations_en}
D.~S. Kulyabov, A.~V. Korolkova, L.~A. Sevastianov, Complex numbers for
  relativistic operations (Dec 2021).
\newblock \href {https://doi.org/10.20944/preprints202112.0094.v1}
  {\path{doi:10.20944/preprints202112.0094.v1}}.

\bibitem{kulyabov_2020_jopcs_hyperbolic-numbers_en}
D.~S. Kulyabov, A.~V. Korolkova, M.~N. Gevorkyan, Hyperbolic numbers as
  einstein numbers, Journal of Physics: Conference Series 1557 (2020)
  012027.1--5.
\newblock \href {https://doi.org/10.1088/1742-6596/1557/1/012027}
  {\path{doi:10.1088/1742-6596/1557/1/012027}}.

\bibitem{fjelstad_1986_extending-sr-perplex-numbers_en}
P.~Fjelstad, Extending special relativity via the perplex numbers, American
  Journal of Physics 54~(5) (1986) 416--422.
\newblock \href {https://doi.org/10.1119/1.14605} {\path{doi:10.1119/1.14605}}.

\bibitem{band_1988_sr-perplex-numbers-comments_en}
W.~Band, Comments on extending relativity via the perplex numbers, American
  Journal of Physics 56~(5) (1988) 469--469.
\newblock \href {https://doi.org/10.1119/1.15582} {\path{doi:10.1119/1.15582}}.

\bibitem{rooney_1978_three-types-complex-number_en}
J.~Rooney, On the three types of complex number and planar transformations,
  Environment and Planning B: Planning and Design 5~(1) (1978) 89--99.
\newblock \href {https://doi.org/10.1068/b050089} {\path{doi:10.1068/b050089}}.

\bibitem{rooney_2014_generalised-complex-numbers_en}
J.~Rooney, Generalised complex numbers in mechanics, in: M.~Ceccarelli, V.~A.
  Glazunov (Eds.), Advances on Theory and Practice of Robots and Manipulators,
  Vol.~22 of Mechanisms and Machine Science, Springer International Publishing,
  Cham, 2014, pp. 55--62.
\newblock \href {https://doi.org/10.1007/978-3-319-07058-2_7}
  {\path{doi:10.1007/978-3-319-07058-2_7}}.

\bibitem{kulyabov_2020_ceur-ws_2639_dual-complex_en}
M.~N. Gevorkyan, A.~V. Korolkova, D.~S. Kulyabov,
  \href{http://ceur-ws.org/Vol-2639/paper-13.pdf}{Approaches to the
  implementation of generalized complex numbers in the julia language}, in:
  D.~S. Kulyabov, K.~E. Samouylov, L.~A. Sevastianov (Eds.), Workshop on
  information technology and scientific computing in the framework of the X
  International Conference Information and Telecommunication Technologies and
  Mathematical Modeling of High-Tech Systems (ITTMM-2020), Vol. 2639 of CEUR
  Workshop Proceedings, Aachen, 2020, pp. 141--157.
\newline\urlprefix\url{http://ceur-ws.org/Vol-2639/paper-13.pdf}

\end{thebibliography}


\begin{thebibliography}{10}
\def\selectlanguageifdefined#1{
\expandafter\ifx\csname date#1\endcsname\relax
\else\selectlanguage{#1}\fi}
\providecommand*{\href}[2]{{\small #2}}
\providecommand*{\url}[1]{{\small #1}}
\providecommand*{\BibUrl}[1]{\url{#1}}
\providecommand{\BibAnnote}[1]{}
\providecommand*{\BibEmph}[1]{#1}
\ProvideTextCommandDefault{\cyrdash}{\iflanguage{russian}{\hbox
  to.8em{--\hss--}}{\textemdash}}
\providecommand*{\BibDash}{\ifdim\lastskip>0pt\unskip\nobreak\hskip.2em plus
  0.1em\fi
\cyrdash\hskip.2em plus 0.1em\ignorespaces}
\renewcommand{\newblock}{\ignorespaces}

\bibitem{bezanson_2017_julia-numeric-computing}
\selectlanguageifdefined{english}
\BibEmph{Bezanson~J., Edelman~A., Karpinski~S., Shah~V.~B.} Julia: A fresh
  approach to numerical computing~//
  \href{http://dx.doi.org/10.1137/141000671}{\BibEmph{SIAM Review}}. \BibDash
\newblock 2017. \BibDash jan. \BibDash
\newblock Vol.~59, no.~1. \BibDash
\newblock P.~65--98.

\bibitem{kulyabov_2014_conf_grid_julia}
\selectlanguageifdefined{english}
\BibEmph{Gevorkyan~M.~N., Kulyabov~D.~S., Sevastyanov~L.~A.} Review of Julia
  programming language for scientific computing~// The 6th International
  Conference "Distributed Computing and Grid-technologies in Science and
  Education". \BibDash
\newblock 2014. \BibDash
\newblock P.~27.

\bibitem{oliphant_book_guide-numpy_en}
\selectlanguageifdefined{english}
\BibEmph{Oliphant~T.~E.} Guide to NumPy. \BibDash
\newblock 2nd edition. \BibDash
\newblock CreateSpace Independent Publishing Platform, 2015. \BibDash
\newblock
  ISBN:~\href{http://isbndb.com/search-all.html?kw=978-1517300074}{978-1517300074}.

\bibitem{zappa-nardelli_2018_julia-subtyping_en}
\selectlanguageifdefined{english}
\BibEmph{{Zappa Nardelli}~F., Belyakova~J., Pelenitsyn~A., Chung~B.,
  Bezanson~J., Vitek~J.} Julia subtyping: a rational reconstruction~//
  \href{http://dx.doi.org/10.1145/3276483}{\BibEmph{Proceedings of the ACM on
  Programming Languages}}. \BibDash
\newblock 2018. \BibDash oct. \BibDash
\newblock Vol.~2, no. OOPSLA. \BibDash
\newblock P.~1--27.

\bibitem{driesen_1995_message-dispatch_en}
\selectlanguageifdefined{english}
\BibEmph{Driesen~K., Hölzle~U., Vitek~J.}
  \href{http://dx.doi.org/10.1007/3-540-49538-x_13}{Message Dispatch on
  Pipelined Processors}~// ECOOP'95 --- Object-Oriented Programming, 9th
  European Conference, Åarhus, Denmark, August 7–11, 1995~/ Ed.\ by\ M.~Tokoro,
  R.~Pareschi. \BibDash
\newblock Lecture Notes in Computer Science. Springer Berlin Heidelberg, 1995.
  \BibDash
\newblock 253–282~p. \BibDash
\newblock
  ISBN:~\href{http://isbndb.com/search-all.html?kw=9783540601609}{9783540601609}.

\bibitem{muschevici_2008_multiple-dispatch-practice_en}
\selectlanguageifdefined{english}
\BibEmph{Muschevici~R., Potanin~A., Tempero~E., Noble~J.}
  \href{http://dx.doi.org/10.1145/1449764.1449808}{Multiple dispatch in
  practice}~// OOPSLA'08: Proceedings of the 23rd ACM SIGPLAN conference on
  Object-oriented programming systems languages and applications. \BibDash
\newblock ACM Press, 2008. \BibDash 10. \BibDash
\newblock P.~563–582.

\bibitem{gowda_2022_symbolic-numerics-multiple-dispatch_en}
\selectlanguageifdefined{english}
\BibEmph{Gowda~S., Ma~Y., Cheli~A., Gwóźzdź~M., Shah~V.~B., Edelman~A.,
  Rackauckas~C.} High-Performance Symbolic-Numerics via Multiple Dispatch~//
  \href{http://dx.doi.org/10.1145/3511528.3511535}{\BibEmph{ACM Communications
  in Computer Algebra}}. \BibDash
\newblock 2022. \BibDash jan. \BibDash
\newblock Vol.~55, no.~3. \BibDash
\newblock P.~92–96.

\bibitem{yaglom_complex-numbers-geometry_ru}
\selectlanguageifdefined{russian}
\BibEmph{Яглом~И.~М.} Комплексные числа и их применение в геометрии~//
  Математика, ее преподавание, приложения и история. \BibDash
\newblock 1961. \BibDash
\newblock Т.~6 {из} \BibEmph{Математическое просвещение, сер. 2}. \BibDash
\newblock {С.}~61--106. \BibDash
\newblock {Режим доступа}: \BibUrl{http://mi.mathnet.ru/mp680}.

\bibitem{yaglom_1964_projective-metrics_ru}
\selectlanguageifdefined{russian}
\BibEmph{Яглом~И.~М., Розенфельд~Б.~А., Ясинская~Е.~У.} Проективные метрики~//
  \BibEmph{Успехи математических наук}. \BibDash
\newblock 1964. \BibDash
\newblock Т.~19, {№} 5 (119). \BibDash
\newblock {С.}~51--113.

\bibitem{kulyabov_2021_preprint_complex-numbers-relativistic-operations_en}
\selectlanguageifdefined{english}
\BibEmph{Kulyabov~D.~S., Korolkova~A.~V., Sevastianov~L.~A.} Complex Numbers
  for Relativistic Operations. \BibDash
\newblock 2021. \BibDash Dec.

\bibitem{kulyabov_2020_jopcs_hyperbolic-numbers_en}
\selectlanguageifdefined{english}
\BibEmph{Kulyabov~D.~S., Korolkova~A.~V., Gevorkyan~M.~N.} Hyperbolic numbers
  as Einstein numbers~//
  \href{http://dx.doi.org/10.1088/1742-6596/1557/1/012027}{\BibEmph{Journal of
  Physics: Conference Series}}. \BibDash
\newblock 2020. \BibDash may. \BibDash
\newblock Vol. 1557. \BibDash
\newblock P.~012027.

\bibitem{fjelstad_1986_extending-sr-perplex-numbers_en}
\selectlanguageifdefined{english}
\BibEmph{Fjelstad~P.} Extending special relativity via the perplex numbers~//
  \href{http://dx.doi.org/10.1119/1.14605}{\BibEmph{American Journal of
  Physics}}. \BibDash
\newblock 1986. \BibDash may. \BibDash
\newblock Vol.~54, no.~5. \BibDash
\newblock P.~416--422.

\bibitem{band_1988_sr-perplex-numbers-comments_en}
\selectlanguageifdefined{english}
\BibEmph{Band~W.} Comments on Extending relativity via the perplex numbers~//
  \href{http://dx.doi.org/10.1119/1.15582}{\BibEmph{American Journal of
  Physics}}. \BibDash
\newblock 1988. \BibDash may. \BibDash
\newblock Vol.~56, no.~5. \BibDash
\newblock P.~469--469.

\bibitem{rooney_1978_three-types-complex-number_en}
\selectlanguageifdefined{english}
\BibEmph{Rooney~J.} On the Three Types of Complex Number and Planar
  Transformations~//
  \href{http://dx.doi.org/10.1068/b050089}{\BibEmph{Environment and Planning B:
  Planning and Design}}. \BibDash
\newblock 1978. \BibDash
\newblock Vol.~5, no.~1. \BibDash
\newblock P.~89--99.

\bibitem{rooney_2014_generalised-complex-numbers_en}
\selectlanguageifdefined{english}
\BibEmph{Rooney~J.}
  \href{http://dx.doi.org/10.1007/978-3-319-07058-2_7}{Generalised Complex
  Numbers in Mechanics}~// Advances on Theory and Practice of Robots and
  Manipulators~/ Ed.\ by\ M.~Ceccarelli, V.~A.~Glazunov. \BibDash
\newblock Cham~: Springer International Publishing, 2014. \BibDash
\newblock Vol.~22 of \BibEmph{Mechanisms and Machine Science}. \BibDash
\newblock P.~55--62.

\bibitem{kulyabov_2020_ceur-ws_2639_dual-complex_en}
\selectlanguageifdefined{english}
\BibEmph{Gevorkyan~M.~N., Korolkova~A.~V., Kulyabov~D.~S.} Approaches to the
  implementation of generalized complex numbers in the Julia language~//
  Workshop on information technology and scientific computing in the framework
  of the X International Conference Information and Telecommunication
  Technologies and Mathematical Modeling of High-Tech Systems (ITTMM-2020)~/
  Ed.\ by\ D.~S.~Kulyabov, K.~E.~Samouylov, L.~A.~Sevastianov. \BibDash
\newblock Vol.~2639 of \BibEmph{CEUR Workshop Proceedings}. \BibDash
\newblock Aachen, 2020. \BibDash apr. \BibDash
\newblock P.~141--157. \BibDash
\newblock Access mode: \BibUrl{http://ceur-ws.org/Vol-2639/paper-13.pdf}.

\end{thebibliography}

\end{document}


\graphicspath{{image/multiple-dispatch/ru/}{image/multiple-dispatch/}{image/}}

\title{Реализация гиперболических комплексных чисел на языке Julia}

\author{А. В. Королькова}
\email{korolkova-av@rudn.ru}
\affiliation{
  Российский университет дружбы народов,\\
  117198, Москва, ул. Миклухо-Маклая, д. 6}

\author{М. Н. Геворкян}
\email{gevorkyan-mn@rudn.ru}
\affiliation{
  Российский университет дружбы народов,\\
  117198, Москва, ул. Миклухо-Маклая, д. 6}

\author{Д. С. Кулябов}
\email{kulyabov-ds@rudn.ru}
\affiliation{
  Российский университет дружбы народов,\\
  117198, Москва, ул. Миклухо-Маклая, д. 6}
\affiliation{
  Объединённый институт ядерных исследований,\\
  ул. Жолио-Кюри 6, Дубна, Московская область, Россия, 141980}

\begin{abstract}
\textbf{Предпосылки.}
Гиперболические комплексные числа применяются при описании
гиперболических пространств. Одним из известных примером таких
пространств является пространство Минковского, играющее ведущее
значение в задачах частной теории относительности, электродинамики.
Однако такие числа не очень распространены в разных языках
программирования.
\textbf{Цель.}
Представляет интерес реализация гиперболических комплексных в языках
научного программирования, в частности, в языке Julia.
\textbf{Методы.}
В основе языка Julia лежит концепция множественной
диспетчеризации (multiple dispatch).
Эта концепция является расширением концепции полиморфизма для
объектно-ориентированных языков программирования.
Для реализации гиперболических комплексных чисел использован
подход множественной диспетчеризацию языка Julia.
\textbf{Результаты.}
В результате получена библиотека, реализующая гиперболические числа.
\textbf{Выводы.}
По результатам исследования можно сделать вывод об удобстве и  естественности
концепции множественной диспетчеризации в языках научного программирования.
\end{abstract}

  \keywords{%
    язык программирования Julia,
    множественная диспетчеризация,
    абстрактные типы данных,
    конвертация типов,
    параметрические структуры,
    гиперболические комплексные числа
  }

\maketitle

\section{Введение}

Язык программирования Julia~\cite{bezanson_2017_julia-numeric-computing, kulyabov_2014_conf_grid_julia}  --- это
перспективный язык, предназначенный для научных вычислений.  В настоящий момент язык Julia достиг стабильного состояния. По замыслу разработчиков Julia решает \emph{проблему двух языков}. Данная проблема заключается в том, что для быстрого прототипирования, обработки данных и визуализации используется интерпретируемый динамический язык или математический пакет (Python, Matlab и т.д.), а для интенсивных численных расчётов программу приходится переписывать на компилируемом языке со статической типизацией (C/C++, Fortran).

Иллюстрацию данной проблемы можно увидеть на примере
языка Python, который приобрел широкую
популярность в качестве интерфейсного <<языка-клея>>. На нем было
написано большое количество библиотек-обёрток, которые использовали
Python-код для вызова C/C++ и Fortran функций из предварительно
скомпилированных библиотек. Так, например, известная библиотека
NumPy~\cite{oliphant_book_guide-numpy_en} на 51\% состоит из кода на языке Си и лишь на
47\% из кода на языке Python (оставшиеся проценты делят между собой
C++, Fortran, JavaScript и Unix shell).

Язык Julia совмещает в себе гибкости интерпретируемых языков с динамической типизацией и производительность компилируемых языков со статической типизацией.

Базовая часть языка Julia крайне схожа с другими языками научного программирования поэтому не вызывает трудности при освоении. Однако ядро Julia построено вокруг концепцию \emph{множественной диспетчеризации} (multiple dispatch)~\cite{zappa-nardelli_2018_julia-subtyping_en}, которая редко встречается в других языках. Именно в этом механизме лежит существенное отличие Julia от других языков и его понимание существенно для полноценного использования всех преимуществ Julia.

\subsection{Структура статьи}
\label{sec:structure}

В статье авторы уделили большое внимание 
иллюстрации механизма множественной диспетчеризации и других
механизмов, которые близко с ней связаны.

В первой части статьи мы даем необходимые определения и иллюстрируем
концепцию множественной диспетчеризации на простых примерах,
позволяющих понять синтаксис, связанный с этой частью языка и уловить
суть данного подхода. Во второй части мы приводим пример реализации
гиперболических комплексных чисел на языке Julia. Данный пример
позволяет затронуть не только множественную диспетчеризацию, но
и механизм приведения типов, иерархию абстрактных типов, перегрузку
арифметических операторов и задание пользовательских типов данных.

\section{Множественная диспетчеризация}

\subsection{Общие определения}

\emph{Динамическая диспетчеризация} (dynamic dispatch) --- это
механизм, который позволяет выбрать какую из множества реализаций
полиморфной функции (или оператора) следует вызвать в данном
конкретном случае~\cite{driesen_1995_message-dispatch_en}. При этом выбор той или иной реализации
осуществляется на стадии выполнения программы. \emph{Множественная
  диспетчеризация} основывается на динамической диспетчеризации. В
этом случае выбор реализации полиморфной функции делается исходя из
типа, количества и порядка следования аргументов функции. Таким образом
реализуется полиморфизм времени выполнения (runtime polymorphic
dispatch)~\cite{muschevici_2008_multiple-dispatch-practice_en, gowda_2022_symbolic-numerics-multiple-dispatch_en}. Заметим также, что кроме термина <<множественная
диспетчеризация>>, также употребляется термин \emph{мультиметод}.

Механизм множественной диспетчеризации похож на механизм перегрузки
функций и операторов, реализованный, например, в языке C++. Перегрузка
функций, однако, осуществляется исключительно на стадии компиляции,
тогда как множественная диспетчеризация должна работать также и на
стадии выполнения программы (полиморфизм времени выполнения).

  \subsection{Множественная диспетчеризация в Julia}

Для иллюстрации механизма множественной диспетчеризации приведём
следующий пример кода на языке Julia.

\begin{minted}{julia}
function f(x, y)
  println("Общая реализация")
  return x + y
end

function f(x)
  println("Для одного аргумента")
  return x
end

function f(x::Integer, y::Integer)
  println("Реализация для целых чисел")
  return x + y
end

function f(x::String, y::String)
  println("Реализация для строк")
  return x * " " * y
end

function f(x::Tuple{Int, Int}, y::Tuple{Int, Int})
  println("Реализация для кортежей из двух целочисленных элементов")
  return (x[1], x[2], y[1], y[2])
end
\end{minted}

В данном примере мы создали пять реализаций функции $f$, которые
отличаются друг от друга разными сигнатурами. В терминах языка Julia
это означает, что у одной функции $f$ теперь существует четыре разных
\emph{метода}. В первых двух методах мы не использовали аннотаций
типов, поэтому тип аргументов будет определен или на стадии компиляции
или на стадии выполнения программы (как в интерпретируемых
языках). Стоит также отметит, что Julia использует динамическую
JIT-компиляцию (just-in-time), поэтому стадия компиляции от стадии
выполнения отделена для пользователя не явным образом.

Аргументы трех следующих методов аннотированы типами, поэтому будут
вызываться только в случае совпадения типов с аннотациями. В методе
\verb|f| для строк используется оператор конкатенации \texttt{*}. Выбор
знака умножения \texttt{*} вместо более традиционного знака сложения
\texttt{+} обосновывается создателями языка тем, что конкатенация
строк операция не коммутирующая, поэтому более логично использовать
для нее знак умножения, а не сложения, которым чаще все принято
обозначать коммутирующие операции.

Следующий фрагмент кода иллюстрирует работу множественной
диспетчеризации на стадии компиляции. Макрос \mintinline{julia}{@show}
служит для распечатки имени функции и переданных ей аргументов.
\begin{minted}{julia}
  @show f(2.0, 1)
  @show f(2, 2)
  @show f(0x2, 0x1) # числа в шестнадцатеричной системе
  @show f("Строка", "текста")
  @show f(3)
  @show f([1, 2], [3, 4])
  @show f((1, 2), (3, 4))
\end{minted}

  \begin{itemize}
  \item В первой строке мы передали функции аргументы вещественного
    типа (с плавающей точкой), поэтому был осуществлен вызов общей
    реализации. Так как для чисел с плавающей точкой определен
    оператор \verb|+|, то функция выполнилась успешно и дала
    правильный результат.
  \item Во второй и третей строках были вызваны методы для целых
    чисел. Заметим, что тип \mintinline{julia}{Integer} является
    \emph{абстрактным} типом и включает в себя знаковые и беззнаковые
    целые числа размером от 1 до 16 байт, определённые в ядре
    языка. Числа, записанные в шестнадцатеричной системе счисления
    интерпретируются по умолчанию как беззнаковые целые.
  \item В четвертой строке был вызван метод для строк. В пятой строке
    метод для одного аргумента.
  \item В шестой строке в качестве аргументов переданы два
    массива. Операция \texttt{+} определена для массивов, поэтому
    функция выполнилась без ошибок и вернула поэлементную сумму.
  \item В седьмой строке аргументами функции являются кортежи,
    состоящие из двух целых чисел. Так как нами был определен метод
    для такой комбинации аргументов -- функция отработала корректно.
  \end{itemize}

\begin{minted}{julia}
  Общая реализация
  f(2.0, 1) = 3.0
  Реализация для целых чисел
  f(2, 2) = 4
  Реализация для целых чисел
  f(0x02, 0x01) = 0x03
  Реализация для строк
  f("Строка", "текста") = "Строка текста"
  Для одного аргумента
  f(3) = 3
  Общая реализация
  f([1, 2], [3, 4]) = [4, 6]
  Реализация для кортежей из двух целочисленных элементов
  f((1, 2), (3, 4)) = (1, 2, 3, 4)
\end{minted}

Приведённые примеры корректно сработают и в языках, поддерживающих
перегрузку функций и не демонстрируют специфику динамической
диспетчеризации, так как типы аргументов известны на стадии компиляции
и доступны транслятору.

Для проверки работы именно динамического вызова методов рассмотрим
следующий код:
\begin{minted}{julia}
  print("Введите целое число:")
  # Считываем строку и конвертируем в целый тип
  @show n = parse(Int32, readline())
  if n > 0
    x = 1.2; y = 0.1
  else
    x = 1; y = 2
  end
f(x, y)
\end{minted}
Здесь типы значений переменных \verb|x| и \verb|y| не известны на стадии
компиляции, так как зависят от того, какое число введёт пользователь
во время выполнения программы. Тем не менее,  для случая
целочисленных \verb|x| и \verb|y| вызывается соответствующий метод.

  \section {Гиперболические числа}

Мы будем использовать гиперболические числа для
иллюстрации возможностей множественной диспетчеризации языка Julia,
поэтому ограничимся лишь определением и основными арифметическими
операциями.

\emph{Гиперболические числа}~\cite{yaglom_complex-numbers-geometry_ru, yaglom_1964_projective-metrics_ru, kulyabov_2021_preprint_complex-numbers-relativistic-operations_en, kulyabov_2020_jopcs_hyperbolic-numbers_en}, наряду с
эллиптическими и параболическими числами, являются обобщением
комплексных чисел. Гиперболические числа можно определить следующим
образом:
\begin{equation}
  z = x + \mathrm{j} y,\; \mathrm{j}^2 = 1,\; \mathrm{j}\neq \pm 1.
\end{equation}
Величину $j$ будем называть \emph{гиперболической мнимой единицей}, а величины $x$ и $y$ действительной и мнимой частями соответственно.

Для двух гиперболических чисел $z_1 = x_1 + \mathrm{j}y_1$ и $z_2 =
x_2 + \mathrm{j}y_2$ выполняются следующие арифметические операции.
\begin{description}
  \item [Сложение] $z_1 + z_2 = (x_1 + x_2) + \mathrm{j}(y_1 + y_2)$.
  \item [Умножение] $z_1 z_2 = (x_1 x_2 + y_1 y_2) + \mathrm{j}(x_1 y_2 + x_2 y_1)$.
  \item [Сопряжение] $z^{\ast} = x - \mathrm{j} y$.
  \item [Обратное число] $z^{-1} = \dfrac{x}{x^2 + y^2} - \mathrm{j}\dfrac{y}{x^2 - y^2}$.
  \item [Деление] $\dfrac{z_1}{z_2} = \dfrac{x_1 x_2 - y_1 y_2}{x^2_2 - y^2_2} + \mathrm{j}\dfrac{x_1 y_1 - x_1 y_2}{x^2_2 - y^2_2}$.
\end{description}

Реализация гиперболических чисел во многом аналогична реализации
комплексных. Необходимо перегрузить операторы \verb|+|, \verb|-|,
\verb|*| и \verb|/|, функции извлечения корня, возведения в степень,
элементарные математические функции и т.д. При этом для целей
иллюстрации механизма работы множественной диспетчеризации основной
интерес представляют именно арифметические операции. Это обусловлено
тем, что элементарные функции принимают только один аргумент и для них
достаточно определить только один метод. В случае же арифметических
операторов необходимо предусмотреть комбинации аргументов разных
числовых типов. Так, например, должна иметься возможность сложения
гиперболического числа с целым, рациональны, иррациональным числом,
что автоматически затрагивает не только множественную диспетчеризацию,
но и механизмы приведения типов, иерархию абстрактных типов и
перегрузку конструктора по умолчанию.

Поэтому мы ограничимся примерами реализации именно арифметических
операций и все, не затронув более сложные в математическом плане
вычисления разнообразных элементарных функций от гиперболического
числа.

Отметим, что кроме термина гиперболические числа, в литературе
встречаются также термины: двойные числа, расщепленные комплексные
числа, комплексные числа гиперболического типа (double numbers, split
complex numbers, perplex numbers, hyperbolic
numbers)~\cite{fjelstad_1986_extending-sr-perplex-numbers_en,
  band_1988_sr-perplex-numbers-comments_en,
  yaglom_complex-numbers-geometry_ru,
  rooney_1978_three-types-complex-number_en, rooney_2014_generalised-complex-numbers_en}.

  \section{Реализация гиперболических чисел в Julia}

  \subsection{Объявление структуры данных}

\begin{figure}
  \centering
  \includegraphics[width=0.8\linewidth]{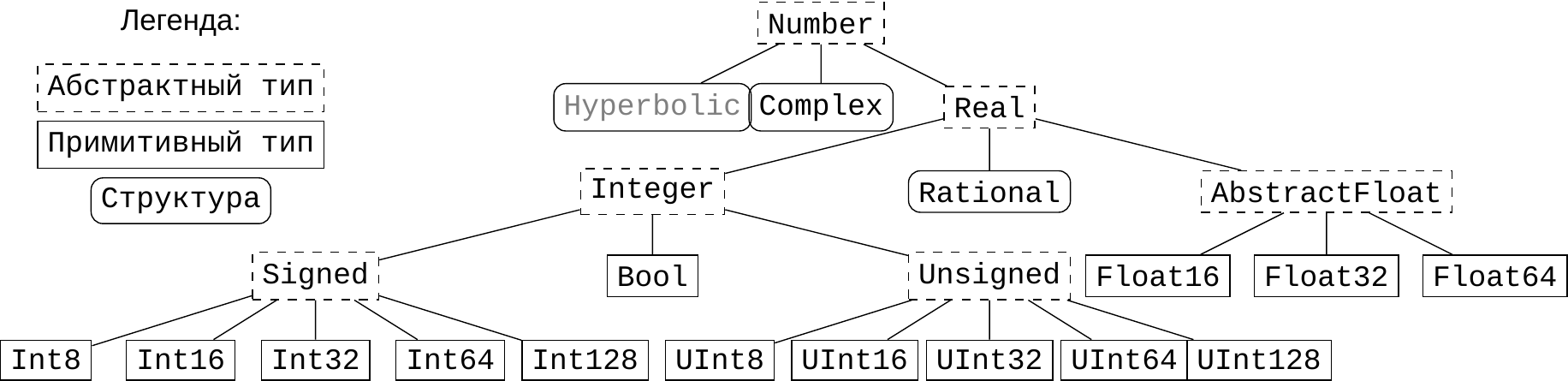}
  \caption{Местоположение гиперболических чисел в иерархии типов Julia}
  \label{fig:01}
\end{figure}

При реализации гиперболических чисел в Julia за основу был взят код
для комплексных чисел, доступный в официальном
репозитории Julia.
Также использовались наработки, полученные при реализации
параболических комплексных чисел~\cite{kulyabov_2020_ceur-ws_2639_dual-complex_en}.
Новый тип \verb|Hyperbolic|
определяется с помощью неизменяемой структуры:
\begin{minted}{julia}
  struct Hyperbolic{T<:Real} <: Number
    "Real part"
    re::T
    "Imaginary part"
    jm::T
  end
\end{minted}

Структура проста и содержит всего два поля параметрического типа
\verb|T|. При этом требуется, чтобы тип \verb|T| был подтипом
абстрактного типа \mintinline{julia}|Real| (синтаксис
\mintinline{julia}|T<:Real|). Сам тип \verb|Hyperbolic| является
подтипом абстрактного типа \mintinline{julia}|Number| (см рис.~\ref{fig:01}). Таким образом
гиперболические числа встраиваются в уже существующую иерархию
числовых типов.

После определения структуры новый объект типа \verb|Hyperbolic| можно
создать путем вызова конструктора по умолчанию. Так, например, число
$h = 1 + \mathrm{j}3$ задается следующим образом:
\begin{minted}{julia}
  h = Hyperbolic{Float64}(1, 3)
\end{minted}

После создания можно обращаться к полям структуры как
\mintinline{julia}{h.re} и \mintinline{julia}{h.jm}, но попытка
изменения значения поля уже существующего объекта приведёт к ошибке,
так как структуры являются неизменяемыми сущностями.

Если оба аргумента конструктора имеют один и тот же тип \verb|T|, то
его можно явно не указывать в фигурных скобках, так как он будет
выведен автоматически из типа передаваемых аргументов.
\begin{minted}{julia}
h = Hyperbolic(1, 3)
\end{minted}

Однако, если типы аргументов отличаются, то конструктор по умолчанию не
сможет осуществить неявное приведение типов и создать новый объект. В
этом случае необходимо явно указывать параметрический тип
\begin{minted}{julia}
  # Float64 и Int64
  h = Hyperbolic(1.0, 3) # Error
  h = Hyperbolic{Float64}(1.0, 3) # Correct
\end{minted}

  \subsection{Дополнительные конструкторы}

Конструктор по умолчанию представляет собой обычную функцию, имя
которой совпадает с именем типа. Создавая дополнительные методы для
этой функции можно создать дополнительные конструкторы для обработки
различных частных случаев.

Так, например, чтобы не указывать всякий раз параметрический тип,
следует добавить новый конструктор следующего вида:
\begin{minted}{julia}
  """Constructor №2"""
  function Hyperbolic(x::Real, y::Real)
    return Hyperbolic(promote(x, y)...)
  end
\end{minted}

Функция \mintinline{julia}{promote} осуществляет приведение типов
переданных ей аргументов к общему типу и возвращает результат в виде
кортежа. Постфиксный оператор \verb|...| распаковывает картеж и
передает его элементы в виде аргументов в функцию-конструктор. В ядре
языка определены правила приведения для всех подтипов абстрактного
типа \mintinline{julia}|Real|, поэтому теперь конструктор будет
корректно работать для любой комбинации аргументов, главное чтобы
выполнялось правило \mintinline{julia}|T<:Real|. Например, следующий
код сработает корректно:
\begin{minted}{julia}
  # Rational и Float64
  h = Hyperbolic(1//3, pi)
  >> Hyperbolic{Float64}(0.5, 3.141592653589793)
\end{minted}

Мы передали в конструктор рациональное число (тип
\mintinline{julia}|Rational|) и встроенную глобальную константу (число
$\pi$) типа \mintinline{julia}|Float64|. После чего сработало правило
приведения типов и оба аргументы были приведены к типу
\mintinline{julia}|Float64| как к более общему.

Объявление еще двух дополнительных конструкторов позволит задавать
гиперболические числа с нулевой мнимой частью:
\begin{minted}{julia}
  """Constructor №3"""
  function Hyperbolic{T}(x::Real) where {T<:Real}
    return Hyperbolic{T}(x, 0)
  end
  """Constructor №4"""
  function Hyperbolic(x::Real)
    return Hyperbolic(promote(x, 0)...)
  end
\end{minted}

Конструктор №3 является параметрической функцией, которая объявляется
с использованием конструкции \mintinline{julia}|where|. Параметр
\verb|T| является подтипом абстрактного типа
\mintinline{julia}|Real|. Конструктор №4 работает аналогично
конструктору №2.

Ещё два конструктора позволят передавать в качестве аргумента
конструктора другие гиперболические числа.
\begin{minted}{julia}
  """Constructor №5"""
  function Hyperbolic{T}(h::Hyperbolic) where {T<:Real}
    Hyperbolic{T}(h.re, h.jm)
  end
  """Constructor №6"""
  function Hyperbolic(h::Hyperbolic)
    return Hyperbolic(promote(h.re, h.jm)...)
  end
\end{minted}

Для большего удобства также можно создать отдельную константу для
мнимой единицы $\mathrm{j}$:
\begin{minted}{julia}
  const jm = Hyperbolic(0, 1)
\end{minted}

  \subsection{Вывод данных}

Для возможности распечатывать значения гиперболического типа в
компактном и читаемом виде, следует добавить соответствующие методы
для функции \mintinline{julia}|show| из модуля
\mintinline{julia}|Base|.
\begin{minted}{julia}
  function Base.show(io::IO, h::Hyperbolic)
    print(io, h.re, "+", h.jm, "j")
  end
\end{minted}

Функция \mintinline{julia}|show| используется при распечатке данных в
консоль, в частности ее вызывают функция \mintinline{julia}|println| и
макрос \mintinline{julia}|@show|. В приведенных далее листингах кода и
результатов его работы будет предполагаться, что добавлен метод
\mintinline{julia}|show| для гиперболических чисел.

  \subsection{Приведение типов}

Прежде чем переходить к реализации методов для арифметических операций
с гиперболическими числами, необходимо определить правила приведения
типов. Для этого следует создать новый метод для функции
\mintinline{julia}|promote_rule| из модуля \mintinline{julia}|Base|.
\begin{minted}{julia}
  function Base.promote_rule(::Type{Hyperbolic{T}}, ::Type{S}) where {T<:Real, S<:Real}
    return Hyperbolic{promote_type(T, S)}
  end
  function Base.promote_rule(::Type{Hyperbolic{T}}, ::Type{Hyperbolic{S}}) where {T<:Real, S<:Real}
    return Hyperbolic{promote_type(T, S)}
  end
\end{minted}

В качестве аргументов в \mintinline{julia}|promote_rule| указываются
параметрические типы, которые следует привести к одному объемлющему
типу. В нашем случае это возможно, если один из типов является
подтипом \mintinline{julia}|Real|, тогда объемлющим типом будет тип
\mintinline{julia}|Hyperbolic|.

После добавления методов для \mintinline{julia}|promote_rule|
становится возможным использовать функции \mintinline{julia}|promote|,
\mintinline{julia}|promote_type| и \mintinline{julia}|convert|.
\begin{minted}{julia}
  >>h = Hyperbolic(1 // 2)
  >>promote(h, 1)
  (1//2+0//1j, 1//1+0//1j)
  >>promote_type(Hyperbolic{Int64}, Float32)
  Hyperbolic{Float32}
\end{minted}

Первая функция нам уже знакома. Вторая же позволяет выводить
объемлющий тип не конкретных значений переменных, а самих типов. Тип в
Julia является объектом первого рода (тип
\mintinline{julia}|DataType|) и его можно присваивать другим
переменным, передавать в качестве аргументов функции и т.д.

Функция \mintinline{julia}|convert| позволяет преобразовать тип
конкретного значения, например:
\begin{minted}{julia}
  >>convert(Hyperbolic, 1)
  1+0j
\end{minted}

\subsection{Арифметические операции над гиперболическими числами}

После добавления методов для приведения типов, можно приступить к
добавлению методов для арифметических операций. Особенностью Julia
является реализация арифметических операций не в виде операторов, а в
виде функций. Так, например, корректны следующие вызовы:
\begin{minted}{julia}
>>+(1,2)
3
>>+(1,2,3,4)
10
>>+((i for i in 1:10)...) # числа от 1 до 10
55
\end{minted}

В связи с этим, добавление методов для арифметических операций ничем
не отличается от соответствующего процесса для других функций.

Добавление методов для унарных операций \verb|+| и \verb|-|
осуществляется следующим образом:
 \begin{minted}{julia}
  Base.:+(h::Hyperbolic) = Hyperbolic(+h.re, +h.jm)
  Base.:-(h::Hyperbolic) = Hyperbolic(-h.re, -h.jm)
\end{minted}
Здесь используется сокращенная запись объявления функции.

Аналогично добавляются методы для бинарного сложения, вычитания,
умножения и деления. Приведем здесь код для сложения и умножения.
\begin{minted}{julia}
  # Binary + and *
  function Base.:+(x::Hyperbolic, y::Hyperbolic)
    xx = x.re + y.re
    yy = x.jm + y.jm
    Hyperbolic(xx, yy)
  end
  function Base.:*(x::Hyperbolic, y::Hyperbolic)
    xx = x.re * y.re + x.jm * y.jm
    yy = x.re * y.jm + x.je * y.re
    return Hyperbolic(xx, yy)
  end
\end{minted}

  \section{Заключение}

Мы рассмотрели механизм множественной диспетчеризации, лежащий в
основе языка Julia, на примере реализации гиперболических
чисел. Данный пример позволил затронуть такие понятия языка как
иерархия типов данных, составные типы данных, механизмы приведения
типов, перегрузка функций (создание новых методов для функций в
терминах языка Julia) и т.д.
\def\acknowledgmentsname{Благодарности}

\begin{acknowledgments}

Публикация выполнена при поддержке Программы стратегического
академического лидерства РУДН.

\end{acknowledgments}

  \ifthenelse{\boolean{xetex} \OR \boolean{luatex}}
  {\bibliographystyle{ugost2008l}}
  {\bibliographystyle{gost2008l}}

\bibliography{bib/multiple-dispatch/cite}